# Molecular dynamics simulations of surface modification formations on polycrystalline Cu under high electric fields

**Kristian Kuppart[1], Simon Vigonski[1,2], Alvo Aabloo[1], Flyura Djurabekova[2,3], Vahur Zadin[1,2*]**

[1]Institute of Technology, University of Tartu, Nooruse 1, 50411 Tartu, Estonia

[2]Helsinki Institute of Physics and Department of Physics, PO Box 43 (Pehr Kalms gata 2), FI-00014, University of Helsinki, Finland

[3]National Research Nuclear University MEPhI, Kashirskoye sh. 31, 115409 Moscow, Russia

## Abstract

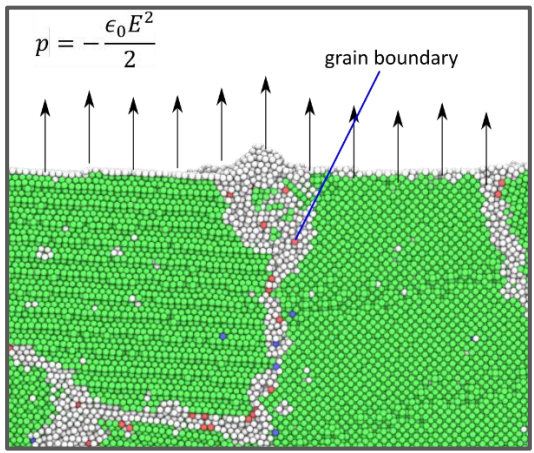

Vacuum breakdowns in particle accelerators and other devices operating at high electric fields is a common problem in the operation of these devices. It has been proposed that the onset of vacuum breakdowns is associated with appearance of surface protrusions while the device is in operation under high electric field. Moreover, the breakdown tolerance of an electrode material was correlated with the type of lattice structure of the material. In the current paper we conduct molecular dynamics simulations of nanocrystalline copper surfaces and show the possibility of protrusion growth under the stress exerted on the surface by an applied electrostatic field. We show the importance of grain boundaries on the protrusion formation and establish a linear relationship between the necessary electrostatic stress for protrusion formation and the temperature of the system. We show that time for protrusion formation increases with the lowering electrostatic field and give the Arrhenius extrapolation to the case of lower fields. General discussion of the protrusion formation mechanisms in the case of polycrystalline copper surfaces is presented.

**Highlights:**

- **Surface grain boundaries promote formation of surface protrusions under applied electric field**
- **Stress needed for protrusion growth decreases linearly with temperature**
- **Time needed for protrusion growth increases exponentially with the lowering of stress**
- **Protrusion growth is caused by diffusion along surface grain boundaries**

---

[*]Corresponding author: vahur.zadin@ut.ee

# 1 Introduction

The planned Compact Linear Accelerator (CLIC) [1] in CERN uses extremely high electric fields (~100 MV/m) to accelerate electrons and positrons to energies up to 3 TeV. Operation under such conditions leads to vacuum breakdowns in the accelerating structures [2], decreasing the luminosity of the collider and increasing the power loss. Moreover, the breakdowns modifying the surface morphology has an overall detrimental effect on the high-precision metal surfaces of the accelerating structures.

Developing a method to reduce the frequency of vacuum breakdowns requires understanding of origins of the phenomenon. Emission current measurements in alternating current [3] and direct current [4], [5] regimes point to the existence of significant field enhancement in the range of $\beta \approx 30 - 150$, the physical origins of which is not entirely clear. It has been theorized [6] that the source for this field enhancement is rough static features residing on the electrode surface acting as field emitters, although surface features with aspect ratio high enough to cause field enhancement of this magnitude have never been observed [5]. Other surface features, such as contaminants, oxides and residual gas are believed to be removed from the electrodes subject to the conditioning process[4] in CLIC and thus not crucial to the breakdown phenomenon. One additional partial explanation for this field enhancement can be given by Schottky's conjecture – the resulting field enhancement from a field emitter imposed on top of another emitter is approximately the product of the respective field enhancements, given that the dimensions of the emitters differ by approximately one order of magnitude [7] [8].

While static field emitters on the surface have not been observed, it has been hypothesized, that these field emitters can dynamically appear by enhanced dislocation activity [9] [1]and surface self-diffusion activity due to the applied electrostatic stress, Joule heating due to emission currents [10], electromigration and other phenomena [11] [12], leading to a self-reinforcing process.

Studies of different materials show that there is a clear correlation between breakdown field strength and crystal structure [2]. Additionally it has been found [5], that the quantity $\beta E$ depends only on the electrode material, where $E$ is the macroscopic mean electric field. Thus, a hypothesis can be formulated that the onset of breakdown is microscopic in nature and that studying the atomistic and crystallographic behavior of materials under high electric fields is a worthwhile endeavor. Previously it has been experimentally shown [13] that nanoscopic ridges can form on the metal electrode from an initial instability due to electric field enhanced surface diffusion. Additionally it has been found that the diffusion coefficient at grain boundaries is much higher than in the bulk of the crystal [14] [15].

Because of the difficulty of conducting atomic level microscopy experiments in high field conditions, computer simulations become a viable means of investigating the accelerator materials. The role of dislocations in surface modification has been studied in simulations of monocrystalline materials. Because of the high stress required to create dislocations in a perfect crystal, various imperfections have been introduced into the simulations. Simulations of subsurface voids [1] and precipitates [16] have shown that these sites can generate plateaus or protrusions on the surface when sufficient external stress is applied.

Nanocrystalline materials have also been studied by simulations. For example, simulations of bulk nanocrystalline copper [17][18] have shown that the microscopic plastic deformation mechanism changes from dislocation based mechanism to grain boundary sliding at grain sizes of $10 - 15$ nm where the yield strength is maximal. At grain sizes in the grain boundary sliding regime, most of new dislocations are generated at the grain boundaries. The effect of tensile stress in copper nanocrystal parallel to the surface has been explored in [19]. It was found, that the increasing stress increases the surface roughness considerably and gives rise to the appearance of surface protrusions.

In this paper, we use molecular dynamics (MD) simulations to study the effect of external electric field on the surface morphology of nanocrystalline copper. MD simulations can provide a detailed microscopic description of the atomistic behavior of the material and its defects. We investigate possible surface protrusion growth mechanisms due to enhanced mobility of atoms on surface grain boundaries. Protrusion growth is analyzed under different conditions of stress due to electrostatic fields and temperature and the protrusion formation time under normal CLIC operating conditions is estimated.

## 2 Methods

The MD simulations were conducted using classical MD code LAMMPS [20]. An Embedded Atom Method (EAM) potential [21] by Mishin *et al.* [22] is used to model the Cu interatomic forces, which has shown good properties for simulating surfaces. In the EAM model, the potential energy of the system is given by:

$$E = \frac{1}{2}\sum_{ij} V(r_{ij}) + \sum_{i} F(\rho_i) \tag{1}$$

Where $V(r_{ij})$ is the pair-potential term and $F(\rho_i)$ is an embedding energy term as a function of the electron density at the atom site $i$ stemming from neighboring atoms.

The velocity-Verlet algorithm with a timestep of $\Delta t = 2$ fs was used throughout this work to propagate the positions and velocities of the atoms. Nose-Hoover [23] and Langevin [24] thermostats were used for temperature and pressure control.

The effect of an external electric field is simulated by applying a tensile stress to the material surface. The effect of an electric field on the surface is simulated as Maxwell stress:

$$\sigma = \frac{\varepsilon_0}{2} F^2 \tag{2}$$

Where $F$ is the applied electric field strength normal to the surface and $\varepsilon_0$ is the vacuum permittivity. This model assumes uniform field strength above the surface, which is accurate as long as the local surface curvature is small. The results can be considered valid until significant surface deformation takes place.

Multiplying the stress by the surface area and dividing by the number of surface atoms gives the force acting on a single surface atom:

$$f = \sigma S/N \qquad (3)$$

This force is added in the MD algorithm to the forces stemming from the neighboring atoms. As a result, surface atoms are pulled away from the surface. The surface

Surface deformation is distinguished from uniform surface strain by comparing the maximum z-coordinate of the surface atoms to the mean z-coordinate of the surface atoms:

$$\delta = \max(z) - \text{mean}(z) \qquad (4)$$

This characteristic is further averaged within a time window of 200 timesteps to decrease noise. The resulting difference is a measure of maximal surface deformation. Rapid increase in $\delta$ indicates the onset of a strong surface deformation and the growth of a protrusion on the surface. It also marks the end of the validity of the uniform electrostatic stress approximation. When $\delta$ exceeds 1 nm, the simulation is stopped. We consider the applied stress at that point to be sufficient to induce significant surface deformation and we define it as the critical stress $\sigma_c$.

## 2.1 Classification of surface atoms

In current study, atoms are classified to belong to the surface with the help of coordination analysis. The physical intuition behind using coordination numbers is simple – the surface atoms have, on average, half the neighbors of bulk atoms. To be able to track the surface even in the case of deformed geometries, the neighboring sphere radius should be taken as large as possible. Due to the LAMMPS implementation for calculating the number of neighboring atoms, this radius cannot exceed the cut-off value for the given interatomic potential and is chosen to be 5 Å, based on the considerations presented below.

Next, we determine a threshold number of neighbors within this radius, which is used to classify the surface atoms. This allows us to determine the atoms that are affected by the external applied electric field (or the corresponding applied force as it is implemented in our simulations) at any given time, as well as dynamically track the surface deformation during the simulation by calculating changes in surface atom positions.

To determine the threshold coordination number, three aspects must be considered. Firstly, it is well established [25] that a static electric field only penetrates into first few atomic layers and thus the algorithm has to consider only the top few surface layers. Secondly, the detected surface should not contain any artificial holes. Thirdly, the amount of low coordination atoms in the bulk classified as surface atoms should be minimal. Taking these restrictions into account, the reasonable neighboring sphere radius is set to be 5 Å and the threshold number of neighbors within that sphere is 37. The sensitivity test simulations showed that within a coordination number range of $37 \pm 3$, the critical stress differs by less than 5%, which is comparable to the statistical uncertainty of the simulations. One example of the resulting surface in deformed state is shown in Figure 1. We can see the existence of small number of wrongly identified surface atoms in the bulk. The number of such atoms decreased further at lower temperatures and in the case of undeformed surfaces.

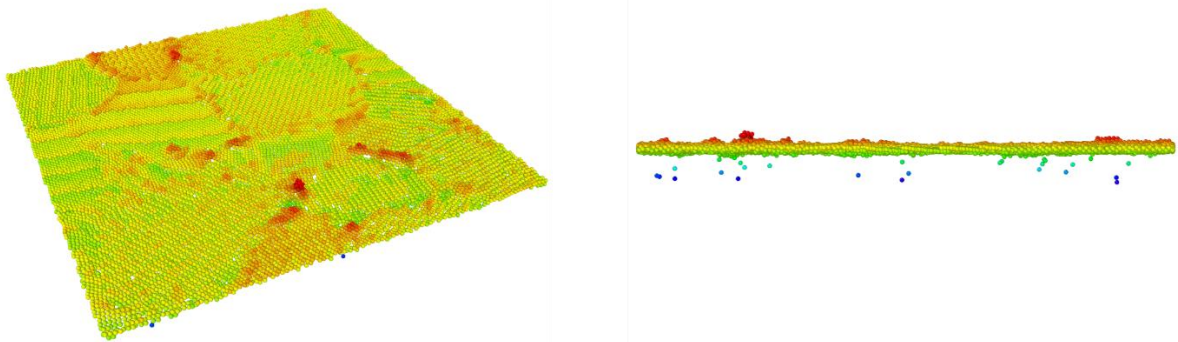

**Figure 1.** Perspective and 2D view of the resulting surface in the case of deformed geometry at 900 K. Atoms are color coded by the z-coordinate. A small number of low coordination false positives can be seen in the rightmost graph.

To give a worst-case, order of magnitude estimate of the effect of the external force on bulk atoms classified as surface atoms, we make the following considerations. First, as the classification of surface atoms is updated every 20 timesteps, we consider the artificial force acting only for this duration. Second, as the number of false positives at temperatures below 900 K is negligible, we use the force calculated from the critical stress at this temperature. Thus we can calculate the additional kinetic energy obtained by an atom in this duration:

$$\Delta E_k = \frac{\Delta p^2}{2m} = \left(\frac{f^2 \Delta t^2}{2m}\right) = 1.2 \text{ meV} \tag{5}$$

Where $f \approx 1$ eV/nm is the force on surface atom corresponding to critical stress value at 900 K, $\Delta t = 40$ fs, and $m$ is the mass of a copper atom. We can compare this energy to the vacancy formation energy for this potential: $E_f = 1.27$ eV. We can see, that this energy is 3 orders of magnitude greater than the additional energy provided by the artificial force. Hence we can conclude that these artifacts cannot influence the simulation outcome.

## 2.2 Generation and preparation of simulated polycrystal

The initial polycrystalline structures were generated with Atomsk [26] using Voronoi tesselation, which has been previously used to create polycrystalline models [17]. 20 different randomized polycrystalline configurations with dimensions of 30 x 30 x 15 nm were created. 10 configurations contained 9 grains, corresponding to mean grain diameter of 13.8 nm, and 10 contained 18 grains, corresponding to a mean grain diameter of 11.0 nm.

Simulations were conducted in two parts, preparation and force ramping. The same simulation procedure was repeated for each of the 20 configurations to obtain a statistical sample of surface processes and to be able to mimic a larger polycrystal configuration. Visualization of atomic configurations and post-processing are done using OVITO [27].

The systems were initially generated as fully periodic. However, since the Voronoi tesselation procedure can lead to structures far from equlibrium, an energy minimization algorithm has to be employed to relax the obtained structures. For that effect, the Polak-Ribiere non-linear conjugate gradient (CG) minimization scheme [28] with relative energy tolerance of $10^{-8}$ was used. The simulation box was allowed to relax possible external pressure components, so that $p_{xx} = p_{yy} = p_{zz} = 0$. An average of 5000 iterative solver steps was required to achieve the required tolerance.

After CG was finished, the systems were considered to be in an equilibrium state for 0 K. To heat the systems up to the temperature used in the simulations, initial velocities of the atoms corresponding to 10K were randomly generated and the temperature was increased using the Nose-Hoover thermostat with a time constant of 0.1 ps. At the same time, Nose-Hoover barostat with a time constant of 1 ps was used to keep zero external pressure to allow for thermal expansion. For each polycrystalline sample, 7 different target temperatures from 300K to 1200K were obtained using such approach. Finally, the box size in the z-direction was increased so that two free surfaces formed and the system is further relaxed for 200 ps with fixed periodic boundaries in the x and y directions at the target temperature, resulting in surface contraction due to the surface tension. As a result of such procedure, the initial states of a polycrystalline copper with open surface were prepared so that surface stress effects were correctly accounted for at given target temperatures. An overview of the preparation and simulation workflow is given in Figure 2.

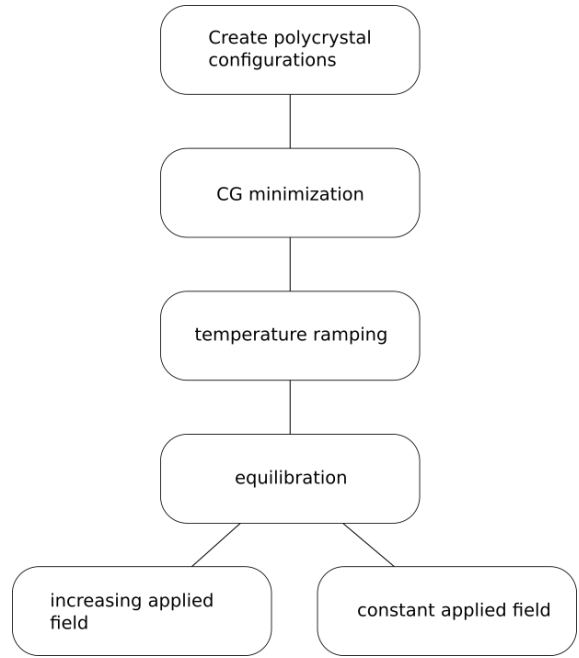

Figure 2: Schematic workflow of the simulation process. 20 polycrystalline structures were obtained after the preparation process. Increasing applied field simulations were conducted for each structure at 7 different temperatures. Constant applied field simulations were conducted for a specific structure at 900 K for 10 fractional values of the critical stress. Each of those were repeated twice to obtain adequate statistics.

## 2.3 Simulations with electric fields

### 2.3.1 Simulations with ramped electric fields

The applied stress was linearly ramped from 0 Pa with a rate of 0.024 GPa/ps. The ramping is necessary to avoid the generation of a strong shockwave through the material due to a sudden application of a high tensile stress. The 3 bottom layers of atoms were fixed by being not included by the integrator, to mimic a semi-infinite bulk material. During ramping, the Langevin thermostat [24] with a time constant of $\gamma = 0.1$ ps was applied to the bulk of the system excluding the surface region, to avoid modifying the relevant dynamics. The Langevin thermostat introduces a source of randomness into the system, which enables statistical sampling of the same configuration with same starting conditions.

### 2.3.2 Simulations with constant applied electric fields

In addition to the simulations with linearly increasing stress, simulations with constant acting stress were also conducted. This simulations were performed with one particular configuration at the temperature of $T = 900$ K. As in the previous case, the temperature was held constant using a Langevin thermostat for all atoms, except the surface and bottom layers. In these simulations, the stress was increased linearly with the same rate as in the previous section, but after the ramping completed, the stress was held constant at a fractional value $\sigma_f$ of the critical stress $\sigma_c$. To account for statistical uncertainty, four simulations with different random number seed were conducted as described in the previous section and the resulting average critical stress of $\sigma_c = 3.07 \pm 0.05$ GPa was obtained. All other boundary and simulation conditions were the same as in the previous section. Simulations with $\frac{\sigma_f}{\sigma_c} = 0.4, 0.45, \dots 0.95$ were conducted and the simulation time needed to reach surface deformation $\delta = 1$ nm was recorded. In addition, a total of 4 simulations were conducted for each value of $\sigma_f$.

## 2.4 Analysis of surface diffusion

### 2.4.1 Temperature dependence of critical mechanical stress

To analyze the temperature dependence of the critical stress, we follow the approach outlined in [29]. We assume that the onset of local surface deformation can be characterized as a thermally activated process, with a mean activation energy per atom $Q(f)$, where $f$ is the force acting on a surface atom in the electric field. We assume, that this activation energy decreases linearly with the force acting on the atom:

$$Q = Q_0 - cf \qquad (6)$$

Where $f$ is the force acting on a surface atom and $c$ is a characteristic length scale corresponding to a particular process. This approach is similar to [30], where the enthalpy of formation of a defect was expressed as:

$$H_f = E_f - \sigma \Delta V \tag{7}$$

Where $E_f$ is the formation energy of the defect, $\sigma = \epsilon_0 E^2/2$ is the Maxwell stress acting on the surface and $\Delta V$ is the relaxation volume of the defect.

The surface diffusion coefficient $D_s$ follows an Arrhenius type equation [29]:

$$D_s = D_0 \exp(-\frac{Q(f)}{kT}) \tag{8}$$

$D_0$ is the exponential pre-factor, which can be calculated as $D_0 = v_0 a^2$. $v_0$ is the attempt frequency on the order of $10^{12} - 10^{13}$s and $a$ is length of an atomic jump, $a \approx 0.3$ nm. We assume that the criterion for the growth of a surface deformation is the surface diffusion coefficient $D_s$ reaching a value corresponding to surface pre-melting, about $D_s = 2 \cdot 10^{-5} \frac{cm^2}{s}$ [29]. Expressing the activation energy in terms of the force acting on the surface atoms and rearranging, we get a linear relationship between the force acting on the surface atoms and the temperature:

$$f = \ln\left(\frac{D_{sc}}{v_0 a^2}\right) \frac{k_b T}{c} + \frac{Q_0}{c} \tag{9}$$

Where $D_{sc}$ is the diffusion constant corresponding to surface pre-melting.

### 2.4.2 Dependence of the deformation time with the force acting on the surface atoms

We assume that the time $\tau_c$ for a surface protrusion to form under constant tensile stress depends exponentially on the activation energy for that process:

$$\tau_c = \tau_0 \exp\left(\frac{Q(f)}{k_b T}\right) \tag{10}$$

Where $t_0$ is the time needed for the protrusion to form in the case of $Q(f) = 0$. We can rewrite the previous equation as:

$$ln\left(\frac{\tau_c}{\tau_0}\right) = \frac{Q_0}{k_b T} - \frac{c}{k_b T} f \tag{11}$$

We can see that there should be a linear relationship between $\ln(\tau_c)$ and the force acting on the surface atoms.

## 3 Results and discussion

### 3.1 Linearly increasing stress

One example of the simulated systems is shown in Figure 3. It can be seen that surface deformation due to the applied electric field starts at grain boundaries or triple junctions lying on the surface, in this particular case, from a quadruple junction. This trend was ubiquitous in all of the ≈ 150 conducted simulations for all simulated configurations and temperatures. Figure 3 (B) shows a height map of the surface at the timestep where a forming protrusion has reached the critical height. Figure 3(A) shows the underlying grain structure in the same location. Atoms are colored using Polyhedral Template Matching [31] to reveal their local lattice type. PTM is an analysis method similar to Common Neighbor Analysis [32], but offers a more robust identification of crystal structure at elevated temperatures [31]. Figure 4 depicts the cross section and the perspective view of the same protrusion. It can be seen from these figures that a single protrusion has formed directly above a quadruple junction. While each configuration had preferential sites for the creation of similar protrusions, the exact location differed from simulation to simulation due to stochasticity of the used Langevin thermostat.

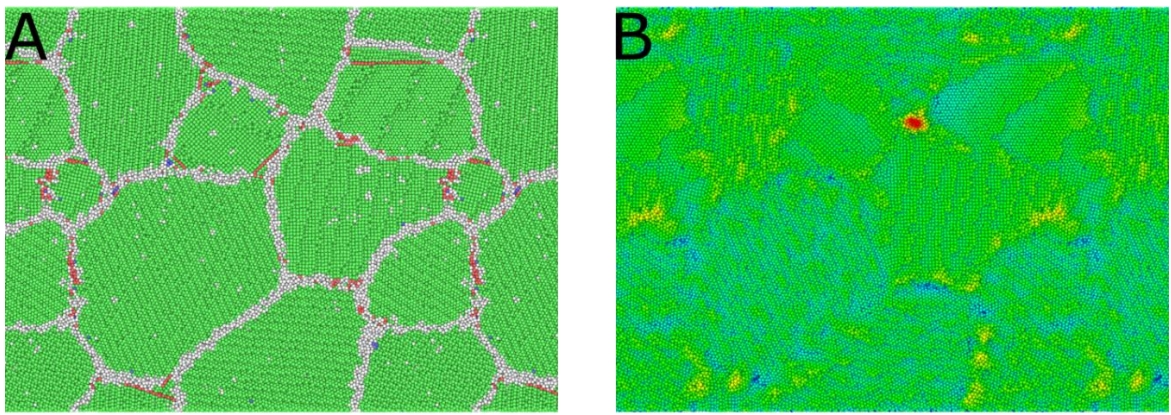

**Figure 3:** A - Grain structure underlying the surface. Atoms are colored by PTM structure type – green: fcc atoms, grey – grain boundary atoms, red: hcp atoms (atoms part of a stacking fault). B – Height map of the surface in the case of critical surface deformation. Atoms are colored by the z-coordinate. Surface protrusion has formed above a quadruple junction.

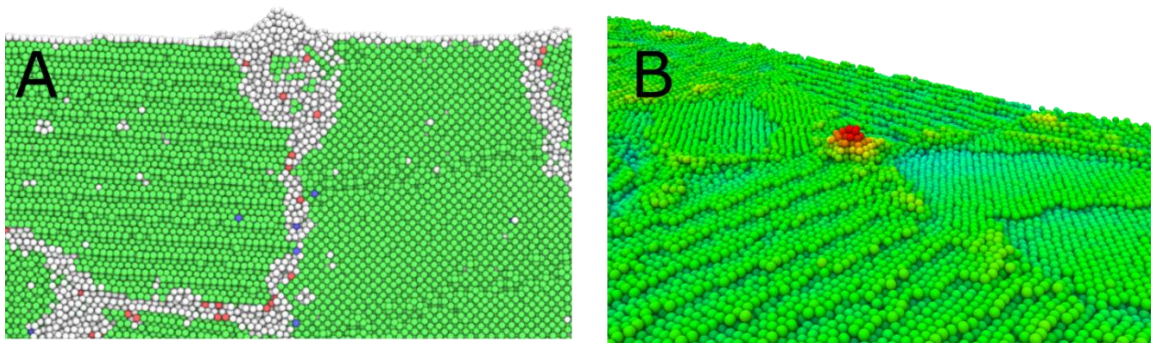

**Figure 4:** The cross section (A) and a perspective view (B) of a surface protrusion. Atoms are colored by the PTM parameter. Video of the protrusion formation is given in supplementary materials.

Figure 5 represents the dependence of the critical stress with temperature. Each data point corresponds to the averaged critical stress over 10 different initial geometries. The error bars represent the standard deviation of the critical stress over different initial geometries. It can be seen that the systems corresponding to these mean diameters do not differ significantly and for that reason, in the following only the cases corresponding to mean grain diameter of 13.8 nm are analyzed. It can be seen from the graph, that to a good approximation, the stress needed for the surface to reach critical deformation decreases linearly with the temperature. The extrapolated value at 0 K is $\sigma_{c0} = 8{,}61$ GPa.

The slope of the line is $-6.1 \cdot 10^{-3} \frac{\text{GPa}}{\text{K}}$ and the x-intercept is $T_c = 1424$ K, slightly higher, than the melting temperature of copper for the used potential [22].

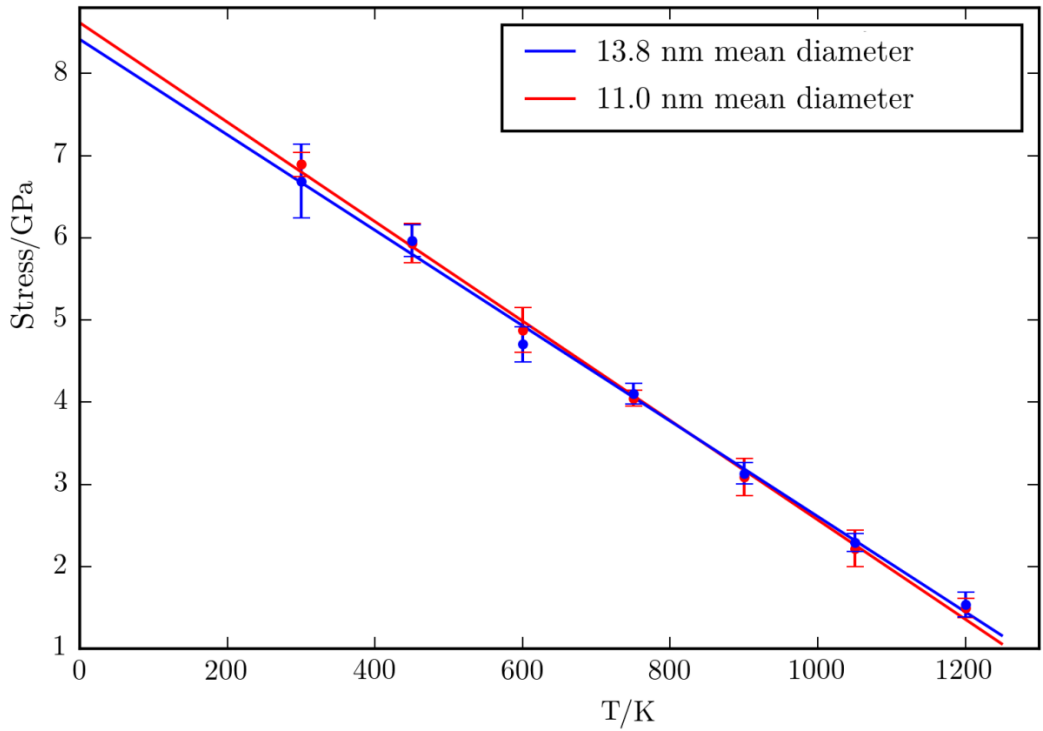

**Figure 5: The dependence of critical stress with temperature**

To follow the approach outlined in 2.4.1, we normalize the stress by the mean number of surface atoms, $N \approx 18500$ and the surface area $A = 30 \text{ nm} \times 30 \text{ nm}$, which is fixed. Doing so we can assess the relevant parameters: $\frac{Q}{c} = 2.6 \frac{\text{eV}}{\text{nm}}, \frac{1}{c} \ln\left(\frac{D_{sc}}{v_0 a^2}\right) = -21 \frac{1}{\text{nm}}$.

It is natural to assume that $c$ is on the order on one atomic spacing, $c \approx 0.3$ nm. Using these results we can calculate the value of the effective activation energy $Q_0 = 0.78$ eV and the surface diffusion coefficient $D_{sc} = 1.65 \cdot 10^{-5} \frac{\text{cm}^2}{\text{s}}$. This activation energy is close to the reported values for the relevant processes, namely surface and grain boundary diffusion [33]. The obtained critical surface diffusion coefficient is close to the reported value for surface pre-melting, which gives some basis for the claim that surface diffusion is the limiting factor in the creation of such surface deformations.

## 3.2 Constant mechanical stress

To analyze the systems in the case of a constant tensile stress acting on the surface, simulations were conducted as described in section 2.3.2. Figure 6 shows the height map of the surface and the underlying grain structure of the particular configuration at the timestep corresponding to the surface deformation of height $\delta_z = 1$ nm electrostatic stress of $\sigma_f = 1.5$ GPa. This stress is significantly lower than the critical stress of the same system at the same temperature, which implies a strong viscoelastic effect in the simulations under study and hints at the possibility that similar deformations would form under even lower stresses, given an evolution time unobtainable by MD simulations. From Figure 6 it can be seen that longer ridges develop along the surface - grain boundary intersections, unlike in the case of linearly increasing stress, where a single surface deformation spot developed. This shows that there are several sites above which a protrusion can form in each configuration.

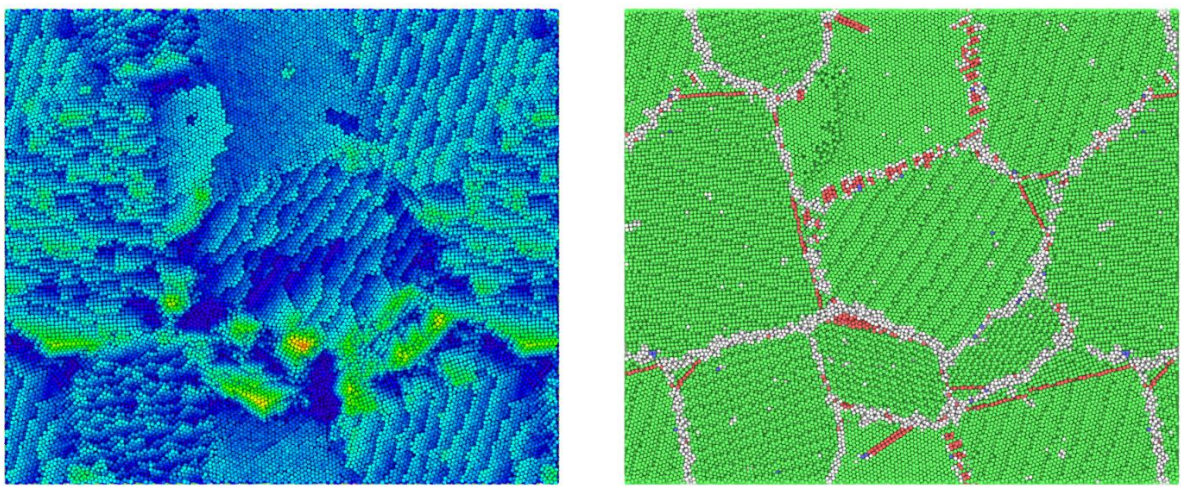

**Figure 6: Height mapping (left) and the underlying grain structure (right) of the system subject to constant tensile stress. A clear tendency of self-roughening is seen under the constant stress, which is lower than critical one in the left image.**

The dependence of the time for a protrusion of height $\delta_z = 1$ nm to form for the given configuration at 900 K is depicted in Figure 7. Each data point corresponds to the average of four simulations. It can be seen from the graph that in the simulated timeframe, critical surface deformation is reached at a surface stress as low as half of the critical stress for that configuration at that temperature.

To quantify the creation of such deformations at a constant tensile stress, we use the results of section 2.4. We estimate the value of $\tau_0$ to be about 3000 timesteps in these simulations. The exact value of this parameter is not crucial, as only the intercept depends on $\ln(\tau_0)$.

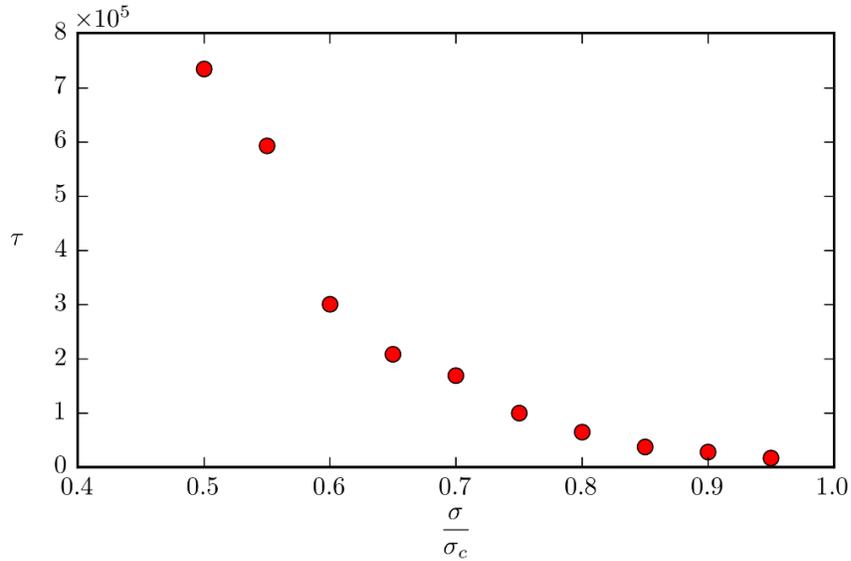

**Figure 7 dependence of the time for a protrusion to form with the force**

Logarithmic plot of the protrusion forming time is given in Figure 8. It can be seen that the graph is linear to a good approximation. From this, we can extract the relevant parameters: $Q_0 = 0.76$ eV, $c_2 = 0.7$ nm. The force proportionality constant is larger than in the previous case, indicating that in the case of constant acting tensile stress and longer simulation times, the stress acting on the surface is more effective in giving rise to surface deformation.

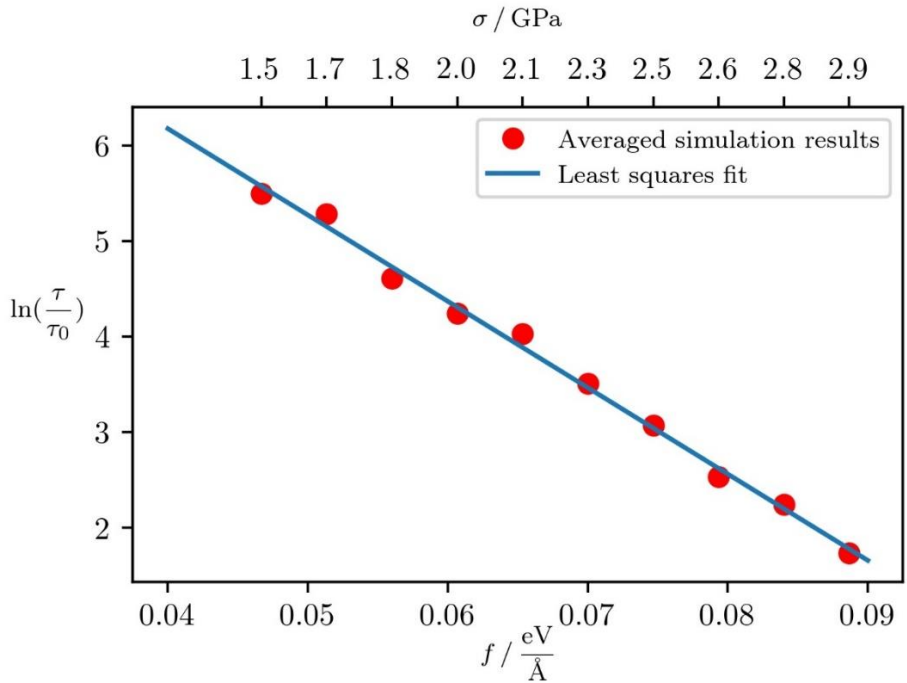

**Figure 8:** Logarithmic graph of the protrusion formation time under different stress conditions. Each Data point was averaged over 3 simulations. Least squares fit used to estimate atomic parameters is given by the blue line.

# 4 Discussion

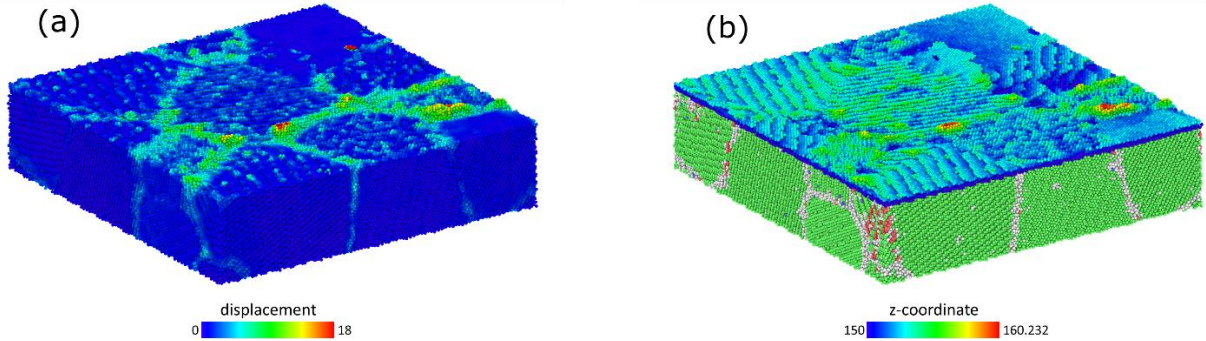

**Figure 9 a) – Total displacement of atoms until critical surface deformation. b) – Height map of the surface and uderlying grain structure. It can be seen that atoms are most mobile on surface and grain boundary intersections. Only a small surface slide is shown for clarity.**

In principle, there are 4 different mechanisms contributing to protrusion formation present in our simulations – surface diffusion, surface grain boundary diffusion, diffusion along bulk grain boundaries and intra-grain dislocation activity. Figure 9 a) shows the total displacement of surface atoms after surface deformation has taken place. From the figure we can qualitatively conclude, that the greatest contribution to protrusion formation stems from diffusion along the intersection of the surface and grain boundaries, also to a lesser extent, along bulk grain boundaries. Rest of the mechanisms seem to have minimal impact on protrusion formation, which allows us to conclude that surface grain boundaries form the primary pathway for surface diffusion. This also explains the results that the protrusion growth preferably starts at grain boundaries and triple junctions.

From the quantities calculated in section 3.2 , we can estimate the time needed for a protrusion to form at CLIC operating conditions of $E = 100 \text{ MV/m}$, $T = 300 \text{ K}$: $\tau \approx 40 \text{ s}$. This time is comparable to the time for the relaxation of similar protrusions as calculated in [34] , which means that even at moderate electric field values there should be a possibility for this kind for surface protrusion formation. These mechanisms are further enhanced due to possible pre-enhancement of the electric field by micro-scale surface irregularities and due to increased temperature from emission currents. While we believe that this simple way of accounting for the effect of electrostatic stress on the surface is valid until the onset of surface deformation, it is possible that atomically rough surface gives rise to a preliminary field enhancement, which accelerates the formation of a protrusion, leading to a self-reinforcing process and further lowers the necessary stress for protrusion formation. Development of models capable of taking into account the field enhancement on the level of single adatoms are left for future work.

# 5 Conclusions

We performed molecular dynamics simulations to simulate the surface deformation and resulting growth of protrusions on nanocrystalline copper surface under electric field. We prepared 20 different nanocrystalline copper configurations equilibrated at 7 different temperatures.

It was found that the protrusions exclusively formed at the intersections of grain boundaries with the surface. This was explained through the increased mobility of surface atoms on grain boundaries, which further underscore the importance of the effect of grain boundaries on the processes on metal surfaces. The electrostatic stress needed to trigger a protrusion formation was found to linearly decrease with the temperature of the system which implies that elevated temperatures, for example from strong local heating from field emission currents, further reinforce the process of such emitter formation. The time for protrusion growth under constant electrostatic stress was found to follow the Arrhenius form, exponentially growing with the lowering of acting stress. The time necessary for a protrusion growth in the low field limit was estimated to be about $\tau \approx 10^2$ s.

# 6 Acknowledgements

This work was supported by Estonian Research Council grants PUT 1372 and IUT 20-24. F. Djurabekova acknowledges gratefully the financial support of Academy of Finland (Grant No. 269696) and MEPhI Academic Excellence Project (Contract No. 02.a03.21.0005). The authors wish to acknowledge High Performance Computation Centre of University of Tartu for providing computational resources.